\newcommand\elph{g}
\newcommand\dos{\rho}
\newcommand\coeff{K}
\newcommand\power{p}
\newcommand\carta{\alpha}
\newcommand\cartb{\beta}
\newcommand\cartc{i}
\newcommand\cartd{j}
\newcommand\banda{n}
\newcommand\bandb{m}
\newcommand\mode{\nu}
\newcommand\intfac{I}
\newcommand\fd{f}
\newcommand\be{n}

\documentclass[aps,prb,reprint,amsmath,noshowpacs,floatfix]{revtex4-1} 

\usepackage{graphicx}

\begin{document}

\title{Accelerated screening of thermoelectric materials by first-principles computations of electron-phonon scattering}

\author{Georgy Samsonidze}
\affiliation{Research and Technology Center, Robert Bosch LLC, Cambridge, MA, USA}
\author{Boris Kozinsky}
\affiliation{Research and Technology Center, Robert Bosch LLC, Cambridge, MA, USA}

\begin{abstract}
Recent discovery of new materials for thermoelectric energy conversion is 
enabled by efficient prediction of materials' performance from first-principles, 
without empirically fitted parameters. 
The novel simplified approach for computing electronic transport properties 
is described, which achieves good accuracy and transferability while greatly 
reducing complexity and computation cost compared to the existing methods. 
The first-principles calculations of the electron-phonon coupling demonstrate 
that the energy dependence of the electron relaxation time varies significantly 
with chemical composition and carrier concentration, suggesting that it is 
necessary to go beyond the commonly used approximations to screen and optimize 
materials' composition, carrier concentration, and microstructure. The new method 
is verified using high accuracy computations and validated with experimental data 
before applying it to screen and discover promising compositions in the space 
of half-Heusler alloys. By analyzing data trends the effective electron mass 
is identified as the single best general descriptor determining material's 
performance. The Lorenz number is computed from first principles and 
the universality of the Wiedemann-Franz law in thermoelectrics is discussed.
\end{abstract}

\date{\today}
\maketitle

\subsection*{Introduction\label{sec:int}}

Discovery of new materials can be greatly accelerated by identifying fundamental physical design rules that govern performance and using them to computationally screen a large number of candidate materials. In the field of thermoelectric (TE) materials this approach is hindered by the difficulty in both measuring and predicting high-temperature transport properties, and their fully first principles prediction has remained out of reach until now. While the electronic and vibrational spectra can be computed using modern ab-initio techniques, the main difficulty lies in being able to describe the interaction of the two spectra which governs much of electronic scattering, thus determining the key transport parameters. In order to connect materials' transport properties to device efficiency, it is useful to compute the TE figure of merit \cite{yan-15-te-descriptor} 
\begin{equation}
ZT = \frac{\sigma S^2 T}{\kappa_e + \kappa_l}
\label{eq:zt}
\end{equation}
where $\sigma$ is the electrical conductivity, $S$ is the Seebeck coefficient, $T$ is the absolute temperature, and $\kappa_e$ and $\kappa_l$ are the electronic and lattice components of the thermal conductivity. Several approaches have been described to compute the lattice thermal conductivity in TE materials from first principles \cite{broido-07-ph-ph-Si-Ge, garg-11-ph-ph-SiGe, carrete-14-hh-lat}, while electronic transport computations so far were limited by computational complexity to simple model systems \cite{restrepo-09-el-ph-Si, bernardi-14-el-ph-Si, qiu-15-el-ph-Si}. In this work we describe a new efficient approach for computing electronic transport properties and present its validation and verification using experiments and more expensive and accurate computations. We then proceed to deploy it to screen materials in a large family of alloys to identify general trends and microscopic factors governing TE performance.

Electronic transport coefficients in semiconductors and metals 
are commonly computed by solving the semiclassical Boltzmann 
transport equation within the relaxation time approximation 
\cite{ziman-64-solid, madsen-06-boltztrap, li-15-el-ph-lerp}. 
In good electrical conductors the electron energy relaxation 
time $\tau$ varies weakly with the electron energy $\epsilon$ 
\cite{ziman-64-solid, singh-97-crt-sk, ahmad-10-tau-model}, 
allowing the use of constant relaxation time 
(CRT) approximation in electronic transport 
calculations \cite{singh-97-crt-sk, blake-01-crt-clathrate, 
scheidemantel-03-crt-BiTe, madsen-06-boltztrap, 
madsen-06-crt-LiZnSb, yang-08-hh, pizzi-14-boltzwann}. 
In this work we show that $\tau$ is not a universal constant and depends
strongly on the material composition and carrier concentration. Moreover,
the $\epsilon$ dependence of $\tau$ can have a significant 
effect on their electronic transport properties. 
This implies that optimizing and screening materials cannot 
be accomplished on the CRT level of approximation, 
and requires the use of first-principles 
computations for predicting $\tau(\epsilon)$.

The total scattering rate $\tau^{-1}$ of electrons 
is approximately the sum of the rates associated 
with intrinsic (electron-electron, electron-phonon) 
and extrinsic (impurities, grain boundaries, alloy disorder) 
scattering mechanisms. In automotive TE power generation, 
the relevant temperature is around $400^\circ$C at the 
hot side of the device, which determines device performance. 
At this temperature, electron-phonon (el-ph) interaction 
is the dominant scattering mechanism, compared to the 
others \cite{bernardi-14-el-ph-Si, yan-15-te-descriptor}. 
First-principles studies of el-ph scattering 
rates in semiconductors and metals have been 
performed using either simplified models, such 
as the deformation potential (DP) approximation 
\cite{sjakste-06-el-ph-GaAs, murphy-armando-08-el-ph-SiGe, 
murphy-armando-10-el-ph-nwSi, wang-11-defpot-Si, 
kaasbjerg-12-defpot, hong-16-defpot-NbFeSb} and Allen's 
formalism \cite{savrasov-96-el-ph, bauer-98-el-ph, xu-14-el-ph-Li}, 
or direct sampling of the el-ph coupling matrix elements 
over the first Brillouin zone (BZ) \cite{sjakste-07-el-ph-GaAs-GaP, 
restrepo-09-el-ph-Si, borysenko-10-el-ph, li-13-el-ph, restrepo-14-el-ph}. 
Due to the high computational cost of direct BZ sampling, 
interpolation schemes were introduced, where 
the couplings are calculated on a coarse grid in the BZ and 
mapped onto a fine grid using linear interpolation 
\cite{li-15-el-ph-lerp} and Wannier interpolation 
(the EPW method) \cite{giustino-07-el-ph, noffsinger-10-epw, 
park-14-el-ph, bernardi-14-el-ph-Si, qiu-15-el-ph-Si, liao-15-el-ph-Si}. 
These methods either do not capture the full details of 
electron scattering or are prohibitively complicated 
to use for realistic materials \cite{yan-15-te-descriptor}. 
For instance, the \texttt{EPW} code \cite{noffsinger-10-epw} 
fully describes the el-ph scattering, but it is challenging 
to apply for materials screening due to the multi-step 
construction procedure for Wannier functions and 
the high computational cost and complexity of the subsequent calculations.

In this work we introduce a new approach, the electron-phonon 
averaged (EPA) approximation, that combines simplicity and speed 
with a fully first-principles treatment of the el-ph interaction. 
It is more predictive than the CRT and DP approximations, 
while allowing for automated rapid calculations for optimization 
of electronic transport quantities, not requiring complex 
interpolation procedures. Its accuracy suggests that energy-dependence 
of scattering is sufficient to quantitatively describe the physics 
of thermoelectricity in complex materials. We first describe 
the approach and validate it using in-depth investigation 
of electron scattering in state of the art TE materials, 
comparing to experiment and the CRT and EPW calculations. 
Then we deploy EPA to perform computational screening 
of the wide half-Heusler (HH) family of compounds 
\cite{wong-ng-13-hh-compositions} for TE power generation applications, 
and examine the material descriptors that determine TE performance 
and can enable high-throughput design.

\subsection*{Theory of electron transport\label{sec:the}}

Solving the semiclassical Boltzmann transport equation for 
electrons within the relaxation time approximation yields 
the following expressions for the electronic transport 
coefficients \cite{ziman-64-solid, madsen-06-boltztrap}: 
\begin{equation}
\sigma_{\carta\cartb}(\mu,T) = \coeff^{(0)}_{\carta\cartb}
\label{eq:sigma}
\end{equation}
\begin{equation}
S_{\carta\cartb}(\mu,T) = k_\mathrm{B} \sum_\cartc 
\big({\coeff^{(0)}}^{-1}\big)_{\cartc\carta} 
\coeff^{(1)}_{\cartc\cartb}
\label{eq:seebeck}
\end{equation}
\begin{equation}
\kappa^e_{\carta\cartb}(\mu,T) = 
k_\mathrm{B}^2 T \Big[\coeff^{(2)}_{\carta\cartb} 
- \sum_{\cartc\cartd} \coeff^{(1)}_{\carta\cartc} 
\big({\coeff^{(0)}}^{-1}\big)_{\cartc\cartd} 
\coeff^{(1)}_{\cartd\cartb}\Big]
\label{eq:kappae}
\end{equation}
where $\carta,\cartb,\cartc,\cartd$ are Cartesian components, 
$\mu$ is the chemical potential of electrons 
(the Fermi level), $T$ is the absolute temperature, 
$k_\mathrm{B}$ is the Boltzmann constant, 
$\sigma_{\carta\cartb}$ is the electrical conductivity, 
$S_{\carta\cartb}$ is the Seebeck coefficient 
(thermopower), $\kappa^e_{\carta\cartb}$ 
is the electronic component of the thermal conductivity, 
and $\coeff^{(\power)}_{\carta\cartb}$ is the 
$\power$-th order electronic transport coefficient. 
The latter is given by: 
\begin{multline}
\coeff^{(\power)}_{\carta\cartb}(\mu,T) = 
\frac{g_s e^{2 - \power}}{(2\pi)^3(k_\mathrm{B} T)^{\power + 1}} 
\sum_{\banda} \int_\mathrm{BZ} d\mathbf{k} 
v_{\banda\mathbf{k}\carta} v_{\banda\mathbf{k}\cartb} \\ 
\times \tau_{\banda\mathbf{k}}(\mu,T) 
\intfac^{(\power)}(\epsilon_{\banda\mathbf{k}},\mu,T)
\label{eq:kepw}
\end{multline}
where $g_s = 2$ is the spin degeneracy, 
$e$ is the elementary charge, 
$\banda$ is the electron band index, 
BZ is the first Brillouin zone, 
$\mathbf{k}$ is the electron wavevector, 
$\mathbf{v}_{\banda\mathbf{k}}$ is the electron group velocity, 
$\tau_{\banda\mathbf{k}}(\mu,T)$ is the electron energy relaxation time, 
$\epsilon_{\banda\mathbf{k}}$ is the electron energy, and 
$\intfac^{(\power)}(\epsilon,\mu,T)$ is the material-independent integrand factor: 
\begin{equation}
\intfac^{(\power)}(\epsilon,\mu,T) = (\epsilon - \mu)^\power 
\fd(\epsilon,\mu,T) \big[1 - \fd(\epsilon,\mu,T)\big]
\label{eq:intfac}
\end{equation} 
Here, $\fd(\epsilon,\mu,T)$ is the Fermi-Dirac distribution 
function. The electron group velocity is defined by: 
\begin{equation}
v_{\banda\mathbf{k}\carta} = \frac{1}{\hbar} 
\frac{\partial\epsilon_{\banda\mathbf{k}}}{\partial k_\carta}
\label{eq:vel}
\end{equation}
where $\hbar$ is the reduced Planck constant, 
and can be evaluated using the Fourier interpolation 
\cite{scheidemantel-03-crt-BiTe, madsen-06-boltztrap} 
or the Wannier interpolation \cite{pizzi-14-boltzwann} 
of $\epsilon_{\banda\mathbf{k}}$.

The inverse of the electron energy relaxation time 
induced by the electron-phonon (el-ph) interaction 
is given by \cite{giustino-07-el-ph, li-15-el-ph-lerp}:
\begin{multline}
\tau_{\banda\mathbf{k}}^{-1}(\mu,T) = 
\frac{\Omega}{(2\pi)^2\hbar} \sum_{\bandb\mode} 
\int_\mathrm{BZ} d\mathbf{q} 
\left|\elph_{\bandb\banda\mode}(\mathbf{k},\mathbf{q})\right|^2 \\ 
\times \bigg\{ 
\Big[\be(\omega_{\mode\mathbf{q}},T) + 
\fd(\epsilon_{\bandb\mathbf{k}+\mathbf{q}},\mu,T)\Big] 
\delta\left(\epsilon_{\banda\mathbf{k}} + \omega_{\mode\mathbf{q}} - 
\epsilon_{\bandb\mathbf{k}+\mathbf{q}}\right) \\ 
+ \Big[\be(\omega_{\mode\mathbf{q}},T) + 1 - 
\fd(\epsilon_{\bandb\mathbf{k}+\mathbf{q}},\mu,T)\Big] 
\delta\left(\epsilon_{\banda\mathbf{k}} - \omega_{\mode\mathbf{q}} - 
\epsilon_{\bandb\mathbf{k}+\mathbf{q}}\right) 
\bigg\} 
\label{eq:epw}
\end{multline}
where $\Omega$ is the volume of the primitive cell, 
$\bandb$ is the electron band index, 
$\mode$ is the phonon mode index, 
$\mathbf{q}$ is the phonon wavevector, 
$\omega_{\mode\mathbf{q}}$ is the phonon energy, 
$\elph_{\bandb\banda\mode}(\mathbf{k},\mathbf{q})$ 
is the el-ph coupling matrix element, 
$\be(\omega,T)$ is the Bose-Einstein distribution 
function, and $\delta$ is the Dirac delta function.

The main element of the EPA approximation is to turn 
the complex momentum-space integration in Eq.~(\ref{eq:epw}) 
into an integration over energies. This is accomplished by 
replacing momentum-dependent quantities in Eq.~(\ref{eq:epw}) 
by their energy-dependent averages. First, el-ph coupling 
matrix elements are averaged over the directions of 
$\mathbf{k}$ and $\mathbf{k}+\mathbf{q}$ wavevectors:
\begin{equation}
\left|\elph_{\bandb\banda\mode}(\mathbf{k},\mathbf{q})\right|^2 
\mapsto \elph_{\mode}^2 
(\epsilon_{\banda\mathbf{k}},\epsilon_{\bandb\mathbf{k+q}}) 
\label{eq:geps} 
\end{equation}
Second, $\mathbf{q}$-dependent phonon energies are 
averaged over the cells of electron energy grids:
\begin{equation}
\omega_{\mode\mathbf{q}} \mapsto \omega_{\mode} 
(\epsilon_{\banda\mathbf{k}},\epsilon_{\bandb\mathbf{k+q}}) 
\label{eq:weps}
\end{equation}
At temperatures below the Debye temperature, 
electron scattering is dominated by acoustic phonons 
\cite{seitz-48-el-ph-acoustic, sjakste-07-el-ph-GaAs-GaP, 
murphy-armando-08-el-ph-SiGe, park-14-el-ph}, 
implying that very fine sampling of electron energies 
is required in Eqs.~(\ref{eq:geps}) and~(\ref{eq:weps}). 
However, in real doped semiconductor samples, 
extrinsic scattering mechanisms are often dominant 
at low temperatures. At temperatures above the Debye 
temperature all the phonon modes are populated, which 
allows using much coarser sampling of electron energies 
in Eqs.~(\ref{eq:geps}) and~(\ref{eq:weps}) to reduce 
computational cost. If electron energy grid spacing is larger 
than the highest optical phonon energy, Eq.~(\ref{eq:weps}) 
is reduced to the average phonon mode energy: 
\begin{equation}
\omega_{\mode\mathbf{q}} \mapsto \bar{\omega}_{\mode} 
\label{eq:wavg}
\end{equation}
This allows further simplification by performing the 
integration over $\mathbf{q}$ and the summation over 
$\bandb$ in Eq.~(\ref{eq:epw}) analytically, which yields:
\begin{multline}
\tau^{-1}(\epsilon,\mu,T) = \frac{2\pi\Omega}{g_s\hbar} \sum_{\mode} \\ 
\bigg\{ 
\elph_{\mode}^2(\epsilon,\epsilon+\bar{\omega}_{\mode}) 
\Big[\be(\bar{\omega}_{\mode},T) + \fd(\epsilon + \bar{\omega}_{\mode}, \mu, T)\Big] 
\dos\left(\epsilon + \bar{\omega}_{\mode}\right) \\ 
+\elph_{\mode}^2(\epsilon,\epsilon-\bar{\omega}_{\mode}) 
\Big[\be(\bar{\omega}_{\mode},T) + 1 - \fd(\epsilon - \bar{\omega}_{\mode}, \mu, T)\Big] 
\dos\left(\epsilon - \bar{\omega}_{\mode}\right) 
\bigg\} 
\label{eq:epa}
\end{multline}
Here, $\dos(\epsilon)$ is the electron density 
of states defined as the number of electronic 
states per unit energy and unit volume. Consequently, 
Eq.~(\ref{eq:kepw}) is rewritten in the following form:
\begin{multline}
\coeff^{(\power)}_{\carta\cartb}(\mu,T) = 
\frac{g_s e^{2 - \power}}{(2\pi)^3(k_\mathrm{B} T)^{\power + 1}} 
\int d\epsilon v_{\carta\cartb}^2(\epsilon) \dos\left(\epsilon\right) \\ 
\times \tau(\epsilon,\mu,T) \intfac^{(\power)}(\epsilon,\mu,T)
\label{eq:kepa}
\end{multline}
where $v_{\carta\cartb}^2(\epsilon)$ is the energy 
projected squared velocity tensor \cite{madsen-06-boltztrap}: 
\begin{equation}
v_{\carta\cartb}^2(\epsilon) \dos\left(\epsilon\right) = 
\sum_{\banda} \int_\mathrm{BZ} d\mathbf{k} 
v_{\banda\mathbf{k}\carta} v_{\banda\mathbf{k}\cartb} 
\delta(\epsilon - \epsilon_{\banda\mathbf{k}})
\label{eq:v2epa}
\end{equation}

\subsection*{Carrier scattering and transport\label{sec:tra}}

To verify our approximation and to compare available approaches, 
we use the CRT, EPA, and EPW methods to compute electronic transport 
properties for state-of-the-art TE materials from the family of HH compounds, 
the $p$-type \mbox{HfCoSb} \cite{culp-08-hh-p-type, yan-12-hh-p-type} 
and $n$-type \mbox{HfNiSn} \cite{joshi-11-hh-n-type}. 
Isovalent Zr-Hf alloying is used to reduce lattice thermal conductivity by 
mass disorder scattering \cite{culp-08-hh-p-type, yan-12-hh-p-type, joshi-11-hh-n-type}, 
and we neglect its effect in our computations of the electronic structure. 
Heterovalent Sn-Sb substitution is experimentally used to achieve $p$- and $n$-doping, 
which we treat within the rigid-band approximation \cite{grimvall-72-rigid-band}. 
The carrier concentrations are obtained from 
Hall measurements at room temperature: 
$p = 0.06$ per formula unit ($1.1 \times 10^{21}$ cm$^{-3}$) for 
\mbox{Hf$_{0.5}$Zr$_{0.5}$CoSb$_{0.8}$Sn$_{0.2}$} \cite{culp-08-hh-p-type} 
and $n = 0.01$ per formula unit ($1.7 \times 10^{20}$ cm$^{-3}$) for 
\mbox{ZrNiSn$_{0.99}$Sb$_{0.01}$} \cite{xie-14-hh-n-type}.

\begin{figure}
\includegraphics[width=\columnwidth]{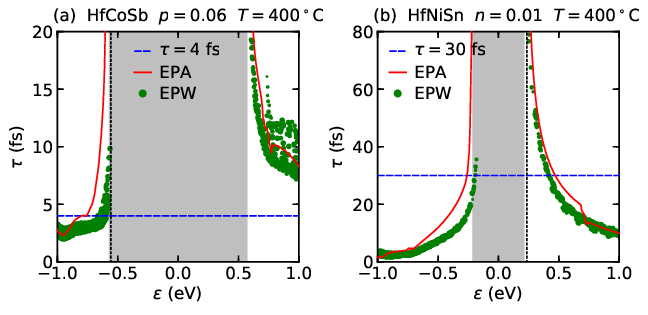}
\caption{The electron energy relaxation time $\tau$ 
for $p$-type \mbox{HfCoSb} and $n$-type \mbox{HfNiSn} 
as a function of the electron energy $\epsilon$ 
calculated within the EPA approximation 
(solid curves) and the EPW method (solid symbols). 
Grey bars denote the band gaps. 
Calculations are performed at temperature $T = 400^\circ$C 
and at carrier concentrations $p = 0.06$ and $n = 0.01$ per 
formula unit for \mbox{HfCoSb} and \mbox{HfNiSn}, respectively. 
The vertical short-dashed lines denote the 
corresponding chemical potentials of electrons $\mu$.
\label{fig:tau}}
\end{figure}

We compare the EPA and EPW results for the 
electron energy relaxation time $\tau$ as a function 
of the electron energy $\epsilon$ on Fig.~\ref{fig:tau} 
and see that the two approaches are in quantitative 
agreement and that $\tau$ increases sharply near 
the band edges. This can be seen by extracting 
the $\epsilon$ dependence from Eq.~(\ref{eq:epa}): 
\begin{equation}
\tau^{-1}(\epsilon) \sim \elph^2(\epsilon) \dos(\epsilon) 
\label{eq:tau} 
\end{equation}
Given a weaker $\epsilon$ dependence of $\elph$ 
than that of $\dos$ (see Supplementary Fig.~1), 
Eq.~(\ref{eq:tau}) implies that $\tau$ varies 
inversely as $\dos$, in agreement with previous 
studies \cite{ravich-70-chalcogenides, mahan-96-best-te, 
yan-15-te-descriptor, fischetti-88-el-tran, 
bernardi-14-el-ph-Si, xu-14-el-ph-Li}. 
This allows $\elph^2$ to be calculated on a coarse 
grid in the BZ and mapped onto a coarse energy 
grid, while $\dos$ is easily computed on a fine grid. 
As a result, the EPA approximation is computationally 
faster by orders of magnitude and simpler to apply 
than the direct BZ sampling and the EPW interpolation 
scheme (see Methods section).

\begin{figure}
\includegraphics[width=\columnwidth]{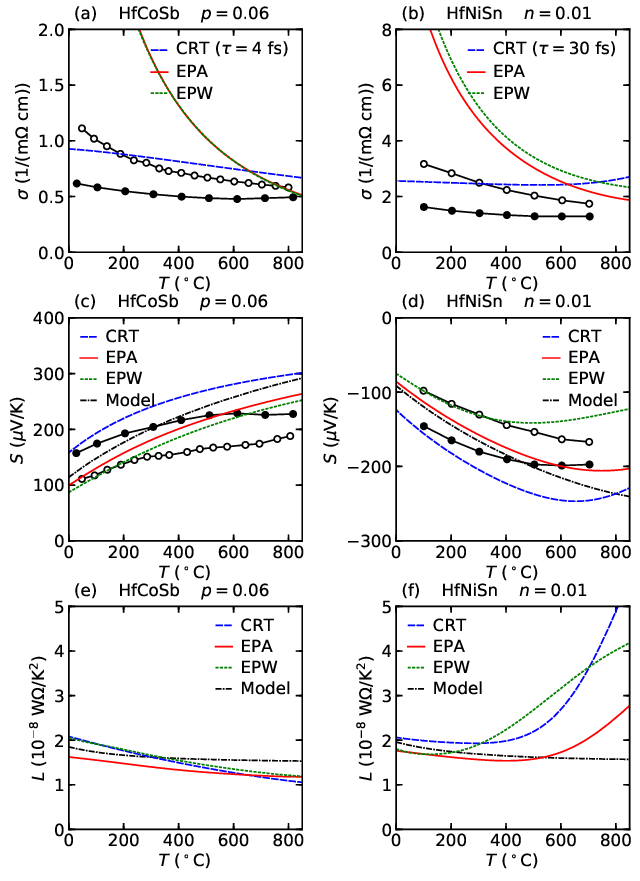}
\caption{The electrical conductivity $\sigma$, 
the Seebeck coefficient $S$, and the Lorenz number $L$ 
for $p$-type \mbox{HfCoSb} and $n$-type \mbox{HfNiSn} 
as a function of temperature $T$ calculated within 
the CRT approximation (long-dashed curves), 
the EPA approximation (solid curves), 
and the EPW method (short-dashed curves). 
Calculations are performed at carrier concentrations 
$p = 0.06$ and $n = 0.01$ per formula unit. 
The dash-dot curves are obtained using 
a single parabolic band model with acoustic 
phonon scattering \cite{fistul-1969-heav-dop-sc}. 
The open symbols show the experimental data for ingot samples 
of $p$-type \mbox{Hf$_{0.5}$Zr$_{0.5}$CoSb$_{0.8}$Sn$_{0.2}$} 
\cite{culp-08-hh-p-type} and $n$-type 
\mbox{Hf$_{0.75}$Zr$_{0.25}$NiSn$_{0.99}$Sb$_{0.01}$} 
\cite{joshi-11-hh-n-type}. The filled symbols show 
the experimental data for nanostructured samples of 
$p$-type \mbox{Hf$_{0.8}$Zr$_{0.2}$CoSb$_{0.8}$Sn$_{0.2}$} 
\cite{yan-12-hh-p-type} and $n$-type 
\mbox{Hf$_{0.75}$Zr$_{0.25}$NiSn$_{0.99}$Sb$_{0.01}$} 
\cite{joshi-11-hh-n-type}. 
\label{fig:temp}}
\end{figure}

\begin{figure}
\includegraphics[width=\columnwidth]{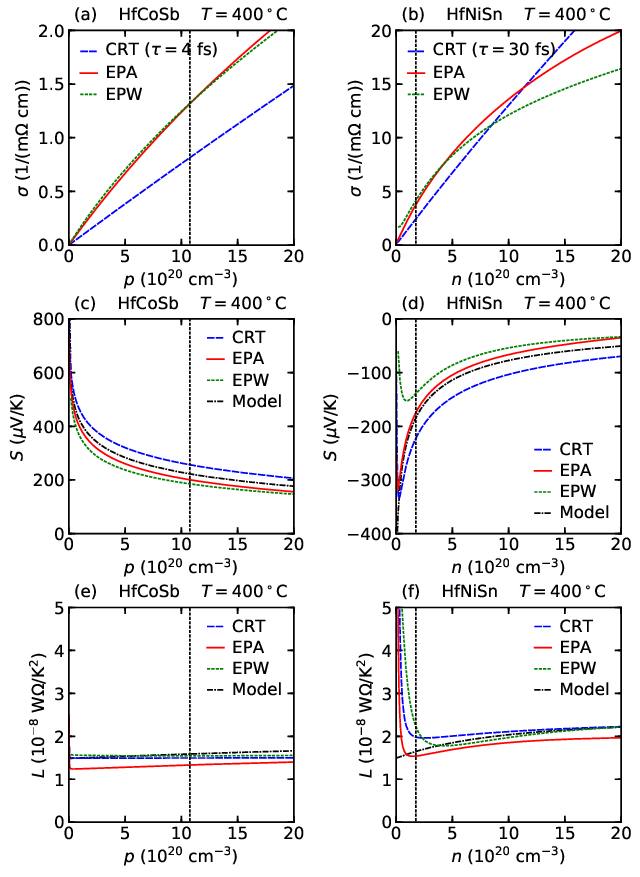}
\caption{The electrical conductivity $\sigma$, 
the Seebeck coefficient $S$, and the Lorenz number $L$ 
for $p$-type \mbox{HfCoSb} and $n$-type \mbox{HfNiSn} 
as a function of carrier concentrations 
$p$ and $n$ calculated within 
the CRT approximation (long-dashed curves), 
the EPA approximation (solid curves), 
and the EPW method (short-dashed curves). 
Calculations are performed at temperature $T = 400^\circ$C. 
The dash-dot curves are obtained using 
a single parabolic band model with acoustic 
phonon scattering \cite{fistul-1969-heav-dop-sc}. 
The vertical short-dashed lines denote 
$p = 0.06$ and $n = 0.01$ per formula unit 
for \mbox{HfCoSb} and \mbox{HfNiSn}, respectively.
\label{fig:x}}
\end{figure}

It is important to note that the values of 
$\tau(\epsilon)$ that contribute the most to electronic transport coefficients, 
particularly conductivities, are selected by the corresponding transport 
integrands (Supplementary Fig.~2) centered at $\mu$, the electronic chemical 
potential that depends on the doping level (vertical lines in Fig.~\ref{fig:tau}). 
The electronic transport coefficients calculated with 
Eqs.~(\ref{eq:sigma})--(\ref{eq:kappae}) are shown in Fig.~\ref{fig:temp} 
as a function of $T$ and in Fig.~\ref{fig:x} as a function of $p$ and $n$. 
For comparison, the experimental data 
for single-crystal ingot and nanostructured samples 
\cite{culp-08-hh-p-type, yan-12-hh-p-type, joshi-11-hh-n-type} 
are shown in Fig.~\ref{fig:temp} by open and filled symbols, respectively. 
The constant values of the electron energy relaxation time, 
$\tau = 4$ fs for heavily doped \mbox{HfCoSb} ($p = 0.06$) and 
$\tau = 30$ fs for less heavily doped \mbox{HfNiSn} ($n = 0.01$), 
are obtained by fitting the calculated electrical conductivity 
$\sigma$ to the experimental data of the ingot samples at 
$T = 400^\circ$C (see Figs.~\ref{fig:temp}(a) and~\ref{fig:temp}(b)). 
The order of magnitude difference between the two 
$\tau$ values indicates that $\tau$ is sensitive 
to both the material composition and the carrier concentration. 
The ability to capture this dependence from first principles, 
which is absent in the CRT approach, is crucial for 
quantitative optimization and screening of materials.

Compared to measurements, EPA and EPW calculations overestimate $\sigma$ at low temperatures, as seen in Figs.~\ref{fig:temp}(a) and~\ref{fig:temp}(b). This is caused by extrinsic scattering mechanisms (not included in our calculations) that decrease the overall $\tau$ at low temperatures. Agreement is much better at higher temperature where el-ph scattering is expected to dominate \cite{bernardi-14-el-ph-Si, yan-15-te-descriptor}. EPA and EPW results agree well for $\tau$ and $\sigma$, which is remarkable given the significant reduction of complexity of the phase-space integral in the EPA scheme. The key implication is that only energy dependence the el-ph scattering is sufficient for accurate description of the physics of electronic relaxation and conductivity.

For the Seebeck coefficient $S$ the agreement between EPA/EPW and experiment is good even in the lower temperature range. This is expected because $S$ is independent of $\tau$ to first order (e.g. on the CRT level of approximation \cite{madsen-06-boltztrap}), and is thus insensitive to extrinsic scattering mechanisms, and determined largely by the electronic band structure. The deviations from experiment at higher temperature are likely due to uncertainties in measured carrier concentrations and computed band structures, as well as their temperature dependence (e.g. thermal expansion, carrier activation). The agreement between EPW and EPA values of $S$ at high $T$ is notably better for \mbox{HfCoSb} than for \mbox{HfNiSn}. This is due to the narrower dispersion in el-ph coupling matrix element values in \mbox{HfCoSb} (see Supplementary Fig.~1), and hence better results from replacing them with averages within the EPA scheme.

The first principles transport formalism provides access also to the electronic part of thermal conductivity $\kappa_e$ (Eq.~(\ref{eq:kappae})), which is important to consider in designing high-performance TE materials, but at the same time is extremely challenging to measure directly by decoupling from the lattice contribution. The Wiedemann-Franz law is commonly used to estimate the $\kappa_e$ from measurements of $\sigma$. The Lorenz number \cite{kumar-93-lorenz} $L = \kappa_e/\left(\sigma T\right)$ is typically set to the Sommerfeld value $L_0=2.44 \times 10^{-8}$ W$\Omega$/K$^2$ (valid for elastic scattering in degenerate electron gas \cite{jones-73-solid, kumar-93-lorenz}), or $L$ can be derived from a single-parabolic-band model with acoustic phonon scattering \cite{fistul-1969-heav-dop-sc, caillat-96-sk, kim-15-lorenz}. To investigate the validity of these approaches we compute $L$ with CRT and, for the first time, using full el-ph intrinsic scattering with EPA and EPW. As Figs.~\ref{fig:temp} and~\ref{fig:x} show, $L$ is far from being constant, deviates significantly from $L_0$, and the single-band model fails to consistently capture its variation. The calculated Lorenz numbers increase as $\mu$ approaches the band edge (Figs.~\ref{fig:x}(e) and~\ref{fig:x}(f)), in agreement with previous observations \cite{hinsche-13-lorenz}. This occurs because the integrand factor of $\kappa_e$ (Eq.~(\ref{eq:intfac})) is broader in energy than the integrand factor of $\sigma$ (see Supplementary Fig.~2), and consequently $\kappa_e$ decreases slower than $\sigma$ as $\mu$ approaches the band edge. The same mechanism causes the well-known increase in $S$ near the band edges in semiconductors (Figs.~\ref{fig:x}(c) and~\ref{fig:x}(d)). Due to the $T$ dependence of $\mu$ and the integrand factors, $L$ can significantly depend on $T$, as is the case for \mbox{HfNiSn}, and less so for \mbox{HfCoSb}. We note that this behavior is primarily a band structure effect and is captured qualitatively already by the CRT calculation. However, we need EPA computations to obtain quantitative values of $\kappa_e$. Importantly, we obtain significant differences between the CRT values of $S$ and $L$ and those computed from EPA and EPW, particularly in the case of \mbox{HfNiSn} (Figs.~\ref{fig:temp} and~\ref{fig:x}), as also observed in other materials \cite{xu-14-el-ph-Li}. The deviation from the CRT results is due to the strong dependence of $\tau$ on $\epsilon$ for \mbox{HfNiSn} in the range of energies where the integrand factors of $S$ and $\kappa_e$ are greatest, as seen by superimposing Supplementary Fig.~2 on Fig.~\ref{fig:tau}. In comparison, the $\tau(\epsilon)$ variation in \mbox{HfCoSb} is an order of magnitude narrower, which leads to better agreement between CRT and EPA/EPW values for $S$ and $L$. The assumption of constant $\tau$ is often used as a justification for the CRT approach for computing $S$ and $L$ in TE materials, where these quantities become independent of the constant $\tau$, hence requiring only the knowledge of the easily computed electronic band structure. We caution that this approach may be simple but is not generally valid in realistic TE materials.

\subsection*{Thermoelectric materials screening\label{sec:scr}}

Having validated the computational approach for well characterized compositions, we turn to the exploration of the wide chemical space of the HH family of alloys. The goal is two-fold: to discover new promising compositions, and to use computed properties to identify broad fundamental design rules that can be used in wider discovery efforts. We start this study by narrowing down the list of all possible elemental combinations in the HH crystal structure, selecting only semiconducting compositions consisting of earth-abundant elements. The final selection of 28 $p$- and $n$-type compounds includes only basic 3-component compositions, expanding upon selections used in previous efforts \cite{wong-ng-13-hh-compositions}. 
Using the EPA formalism we compute the electronic transport 
coefficients $\sigma$, $S$, and $\kappa_e$ for the 28 HH compounds. 
$ZT$ (Eq.~(\ref{eq:zt})) depends also on the lattice contribution to the thermal conductivity 
$\kappa_l$, and there has been work on methodology of computing it from first-principles 
starting from anharmonic phonon scattering \cite{broido-07-ph-ph-Si-Ge, garg-11-ph-ph-SiGe, carrete-14-hh-lat}. 
However, in many TE materials, specifically in HH alloys, 
$\kappa_l$ can be substantially reduced from its intrinsic 
value by isoelectronic alloying and nanostructuring 
\cite{yan-12-hh-p-type, joshi-11-hh-n-type}. This is possible due to
rather short electron mean free paths in these materials, as we show below, and long phonon mean free paths. Using these processing techniques the intrinsic room temperature 
values of $22$ W/(m K) for \mbox{HfCoSb} and $20$ W/(m K) 
for \mbox{HfNiSn} \cite{carrete-14-hh-lat} are reduced 
to $3$ W/(m K) for \mbox{HfCoSb} \cite{yan-12-hh-p-type} 
and $4$ W/(m K) for \mbox{HfNiSn} \cite{joshi-11-hh-n-type}. Consequently, we do not include computations of intrinsic values of $\kappa_l$. As a realistic assumption to explore upper limits of performance, we set $\kappa_l$ for all HH compounds in this work to have the value of $2$ W/(m K) at $T = 400^\circ$C \cite{yan-12-hh-p-type, joshi-11-hh-n-type}.

\begin{figure*}
\includegraphics[width=0.7\textwidth]{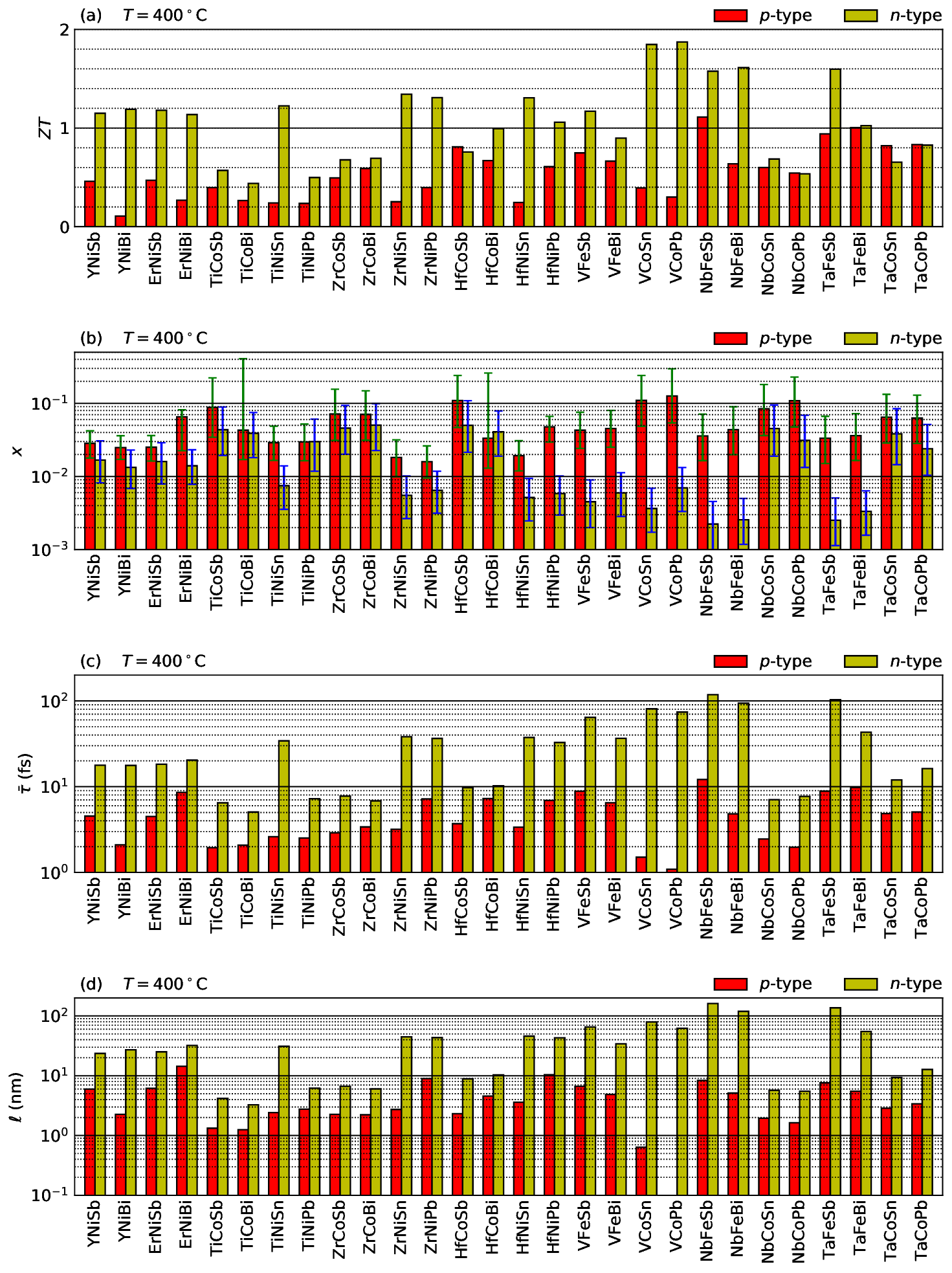}
\caption{The thermoelectric figure of merit $ZT$, 
the optimal carrier concentration $x$ (per formula unit), 
the electron energy relaxation time $\bar{\tau}$, 
and the electron mean free path $\ell$ 
for the $p$- and $n$-type HH compounds 
(shown as red and yellow bars, respectively) 
calculated within the EPA approximation. 
Calculations are performed at temperature 
$T = 400^\circ$C using the lattice thermal 
conductivity $\kappa_l = 2$ W/(m K). 
The values of $x$ are selected to 
maximize the power factor $\sigma S^2$. 
The error bars in (b) show the range where 
$\sigma S^2$ drops by 10\% from its maximum value.
\label{fig:bar}}
\end{figure*}

As mentioned above, electronic transport coefficients depend strongly on 
the electronic chemical potential, and hence the carrier concentration $x$. 
For comparing materials of different compositions we identify the value 
of $x$ for each compound such that the power factor $\sigma S^2$ is maximized. 
We found that there is only 10\% average difference in $ZT$ at values 
of $x$ that maximize $\sigma S^2$ versus those that maximize $ZT$. 
Importantly, the EPA method enables us to perform sweeps in the $x$ values 
automatically and much faster than the EPW procedure (see Methods). 
This step is critical in identifying the maximum achievable performance 
for each composition in the screening. 
The resulting values of $ZT$ and $x$ for the 28 HH compounds 
are shown in Figs.~\ref{fig:bar}(a) and~\ref{fig:bar}(b). 
One can see from Fig.~\ref{fig:bar}(a) that state-of-the-art 
TE materials \mbox{HfCoSb} \cite{yan-12-hh-p-type} and 
\mbox{HfNiSn} \cite{joshi-11-hh-n-type} have some of the highest 
$ZT$ values among $p$- and $n$-type compounds, respectively. 
For these compounds our calculations predict $ZT$ values 
($0.8$ and $1.3$) similar to those determined experimentally 
\cite{yan-12-hh-p-type, joshi-11-hh-n-type} ($0.6$ and $0.8$), 
also at carrier concentrations ($p = 0.11$ and $n = 0.01$) 
similar to the values obtained from the Hall measurements 
\cite{culp-08-hh-p-type, xie-14-hh-n-type} ($p = 0.06$ and $n = 0.01$). 
The best $p$-type compound according to Fig.~\ref{fig:bar}(a) is \mbox{NbFeSb}. We note that we used EPA to computationally identify this composition and suggest for experimental synthesis. A variant of this material was recently synthesized and tested in devices, 
showing leading TE performance as well as material-level cost 
and record thermal cycling reliability \cite{joshi-14-hh}.

Having validated our computational and screening methodology, 
we proceed to analyze the computed dataset in order to distill 
physical trends in the given materials space. 
We first consider the variation of $\tau$ across the space 
of compositions, testing the validity of the common fixed-value 
CRT approximation for materials discovery. 
In order to simplify the comparison between materials, we integrate 
over the energy dependence to define for each compound 
a single effective value for the electronic relaxation time 
$\bar{\tau} = \tau_\mathrm{crt} \sigma_\mathrm{epa} / \sigma_\mathrm{crt}$, 
where $\sigma_\mathrm{epa}$ and $\sigma_\mathrm{crt}$ are electrical 
conductivities computed within the EPA and CRT approximations, 
respectively, and $\tau_\mathrm{crt}$ is the constant relaxation time.
Each quantity is computed at optimal value of $x$ at $T = 400^\circ$C.
Note that $\bar{\tau}$ does not depend on $\tau_\mathrm{crt}$ 
because $\sigma_\mathrm{crt} \propto \tau_\mathrm{crt}$ by definition. 
We also define the effective electron mean free path (MFP), 
$\ell = \bar{\tau} v$, where $v$ is the electron group velocity calculated 
at the value of $\epsilon$ at which $\tau(\epsilon) = \bar{\tau}$. 
The resulting values of $\bar{\tau}$ and $\ell$ for the 28 HH 
compounds are shown in Figs.~\ref{fig:bar}(c) and~\ref{fig:bar}(d). 
The values of $\bar{\tau}$ vary in the range of $1$--$100$ 
fs for different HH compounds (see Fig.~\ref{fig:bar}(c)). 
This implies that a single value of $\tau$ cannot be used for 
different materials. Similarly, the values of $\ell$ vary in the range 
of $1$--$100$ nm for different HH compounds (see Fig.~\ref{fig:bar}(d)). 
Interestingly, these values are too small to be affected by typical 
nanostructuring, given that the average grain size in nanostructured HH 
compounds is greater than $200$ nm \cite{yan-12-hh-p-type, joshi-11-hh-n-type}. 
This explains why the nanostructuring approach is so effective 
for improving TE performance of HH compounds. It does not 
negatively impact $\sigma$ while at the same time it reduces 
$\kappa_l$ \cite{yan-12-hh-p-type, joshi-11-hh-n-type} 
and consequently increases $ZT$. Our findings also 
suggest that decreasing the average grain size below $100$ 
nm may have limited potential for increasing TE performance.

\begin{figure}
\includegraphics[width=\columnwidth]{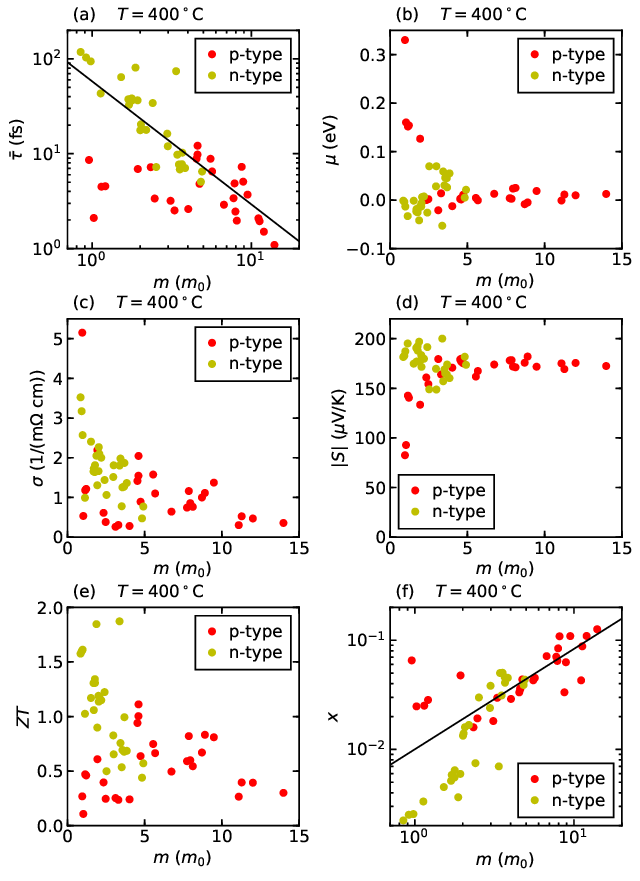}
\caption{The electron energy relaxation time $\bar{\tau}$, 
the chemical potential of electrons $\mu$ (relative to the band edge), 
the electrical conductivity $\sigma$, 
the Seebeck coefficient $S$, 
the thermoelectric figure of merit $ZT$, 
and the carrier concentration $x$ (per formula unit) 
for the $p$- and $n$-type HH compounds 
(shown as red and yellow dots, respectively) 
as a function of the DOS effective mass $m$ 
(in units of free electron mass $m_0$) 
calculated within the EPA approximation. 
Calculations are performed at temperature 
$T = 400^\circ$C using the lattice thermal 
conductivity $\kappa_l = 2$ W/(m K). 
The values of $x$ are selected to 
maximize the power factor $\sigma S^2$. 
The solid lines are power law fits with 
mean exponents and standard deviations of 
(a) $-1.30 \pm 0.26$ and (f) $0.92 \pm 0.09$.
\label{fig:cm}}
\end{figure}

\begin{figure}
\includegraphics[width=\columnwidth]{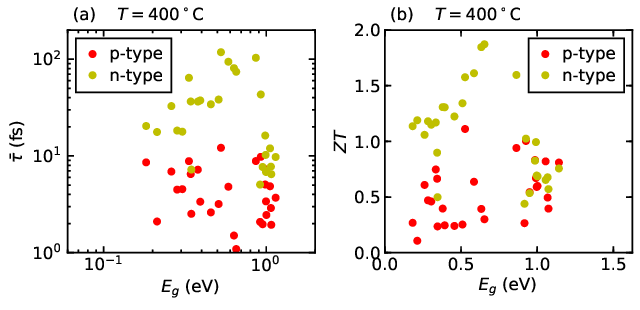}
\caption{The electron energy relaxation time $\bar{\tau}$ 
and the thermoelectric figure of merit $ZT$ 
for the $p$- and $n$-type half-Heusler compounds 
(shown as red and yellow dots, respectively) 
as a function of the band gap $E_g$ 
calculated within the EPA approximation. 
Calculations are performed at temperature 
$T = 400^\circ$C using the lattice thermal 
conductivity $\kappa_l = 2$ W/(m K). 
The values of carrier concentration 
are selected to maximize the power 
factor $\sigma S^2$.
\label{fig:ceg}}
\end{figure}

A computation discovery effort would be greatly facilitated by identifying simple physical descriptors that can be used to predict TE performance of materials. Rapid screening calculations using the EPA method provide a path to this goal by yielding the relationships between electronic and atomic structures and TE properties for a wide set of materials without empirical bias. We analyzed the results of EPA computations on our set of 28 HH compounds by evaluating statistical correlations between transport coefficients at optimal doping and basic characteristics of electron and phonon spectra, such as acoustic phonon velocities, optical phonon frequencies, carrier effective masses (Fig.~\ref{fig:cm}), and electronic band gaps (Fig.~\ref{fig:ceg}). The strongest trend we found is that the electron density-of-states (DOS) effective mass 
$m$ is the single best descriptor of overall performance at optimal doping, as illustrated on Fig.~\ref{fig:cm}. 
We can explain the trends in the transport properties using arguments based on the single parabolic 
band model. In this model the density of states is $\dos \propto m^{3/2} \epsilon^{1/2}$ 
and the electron group velocity is $v \propto m^{-1/2} \epsilon^{1/2}$. Using Eq.~(\ref{eq:tau}) 
we obtain \cite{ravich-70-chalcogenides} $\tau \propto m^{-3/2} \epsilon^{-1/2}$. 
The relevant values of $\epsilon$ are determined by the chemical 
potential of electrons $\mu$ (Fig.~\ref{fig:cm}(b)) and are thus 
mostly independent of $m$ (Fig.~\ref{fig:cm}(b)). The power law 
fit to $m$-dependence of $\bar{\tau}$ (solid line in Fig.~\ref{fig:cm}(a)) indeed gives a close value of $-1.30$ for the exponent. 
The electrical conductivity $\sigma$ (Fig.~\ref{fig:cm}(c)) 
is an integral of $v^2 \dos \tau$ times the 
$\mu$-dependent integrand factor (Eq.~(\ref{eq:kepa})). 
We thus obtain $\sigma \propto m^{-1}$ times an 
$m$-independent integral. Indeed, the envelope of $\sigma$ 
distribution in Fig.~\ref{fig:cm}(c) decreases with $m$. 
The Seebeck coefficient $S$ (Fig.~\ref{fig:cm}(d)) is a ratio of 
two integrals with the same $m$ dependence (Eq.~(\ref{eq:seebeck})). 
Although $S$ increases near the band edge (Fig.~\ref{fig:x}(c)), 
it is independent of $m$ if the band is parabolic. 
Increasing $S$ requires changing the shape of $\dos (\epsilon)$ rather 
than simply increasing $m$ \cite{mahan-96-best-te, heremans-08-PbTe-dos}. 
Accordingly, the $S$ distribution at optimal values of $x$ in Fig.~\ref{fig:cm}(d) shows no overall significant dependence on $m$. Consequently, 
$ZT$ (Fig.~\ref{fig:cm}(e)) carries the $m^{-1}$ dependence 
from $\sigma$. As expected, the envelope of $ZT$ distribution 
in Fig.~\ref{fig:cm}(e) decreases with $m$. The carrier 
concentration $x$ (Fig.~\ref{fig:cm}(f)) is an integral of 
$\dos$ times the $\mu$-dependent Fermi-Dirac distribution function. 
We thus obtain $x \propto m^{3/2}$ times an $m$-independent 
integral. The power law fit (solid line in Fig.~\ref{fig:cm}(f)) 
gives a slightly lower value of $0.92$ for the exponent. 
We can thus rationalize the trends in transport quantities 
in terms of the physics of the simple parabolic band model, even though 
material-specific deviations certainly require accurate computations. 
Based on Fig.~\ref{fig:cm}(e), 
we conclude that the highest $ZT$ values appear 
in the low-$m$ region, which agrees with previous studies that used 
semi-empirical methods \cite{pei-12-low-mass, yan-15-te-descriptor}. 
Note that according to Fig.~\ref{fig:cm}(f), the low-$m$ materials 
generally achieve optimal TE performance at low carrier concentrations.
In summary, we are able to understand and even quantitatively anticipate 
the trends in computed transport properties as a function of the effective 
mass $m$ across a wide set of compositions. Despite the fact that this study was limited to one structural family of half-Heusler alloys, we believe that the trends hold generally for other classes of materials.

\subsection*{Conclusions\label{sec:con}}

We presented a simplified computational method for first-principles prediction of transport properties that achieves good accuracy and transferability at a greatly reduced complexity and computation cost.
Our new approach is suitable for performance optimization and design of next-generation materials for waste hear recovery, exemplified by our computational screening of half-Heusler compounds and identification of a new composition with leading cost, TE performance, and thermal cycling.
We are able to show for the first time from first principles that in TE materials the energy dependence of the electron relaxation time can have a significant effect on their transport properties, including the Seebeck coefficient and the Lorenz number which are generally assumed to be independent of the relaxation time. By directly computing electrical and the electronic part of the thermal conductivities, we find deviations from the Wiedemann-Franz law in these materials at high temperatures and low carrier concentrations. This suggests potential risks in the common procedures used to interpret results of electronic and thermal transport measurements. We demonstrate that the entire complexity of electron-phonon scattering coupling tensor is not needed for accurately calculating electron relaxation times and electronic transport coefficients in TE materials. In addition, we identify the electron effective mass as a useful qualitative descriptor of TE performance, which can be used to screen and prioritize materials. In conclusion, we demonstrate a pathway to wide computational discovery, optimization and understanding of realistic TE materials using first-principles calculations of electronic and vibrational spectra and their coupling. This methodology opens opportunities for understanding intrinsic transport properties of complex semiconductors and enables wide computational materials design in a wide range of technological applications.

\subsection*{Methods\label{sec:met}}

The structural, electronic, and vibrational properties are obtained 
from density functional theory (DFT) \cite{kohn-65-dft} and density 
functional perturbation theory (DFPT) \cite{baroni-01-dfpt} calculations 
using the \texttt{Quantum ESPRESSO} \cite{giannozzi-09-qe} code. 
The electron energy relaxation times and the electronic transport 
coefficients are calculated within the CRT and EPA approximations and 
the EPW method using the \texttt{BoltzTraP} \cite{madsen-06-boltztrap} 
and \texttt{EPW} \cite{noffsinger-10-epw} codes. The EPA method 
is implemented in \texttt{Quantum ESPRESSO} and \texttt{BoltzTraP}. 
Calculations are performed using the generalized gradient approximation 
in the PBE form \cite{perdew-96-pbe} for the exchange-correlation functional, 
ultrasoft pseudopotentials \cite{vanderbilt-90-uspp, dal-corso-14-pslibrary} 
for the core-valence interaction, and a plane wave basis set 
with 80 and 700 Ry kinetic energy cutoffs for wavefunctions 
and charge density. Uniform 8$\times$8$\times$8 $\Gamma$-centered 
$\mathbf{k}$- and $\mathbf{q}$-point grids are used for charge 
density and el-ph calculations, and 48$\times$48$\times$48 grids 
for band structure and transport calculations. The averaging in 
EPA calculations is performed over the cells of an energy grid 
with a spacing of 0.2 eV---the smallest spacing such that all 
cells in the energy grid are filled with $\mathbf{k}$-points (Supplementary Fig.~1). 
Due to incompatibility of the \texttt{EPW} code with ultrasoft 
pseudopotentials and its poor scaling, all calculations 
in Figs.~\ref{fig:tau}, \ref{fig:temp}, and~\ref{fig:x} 
are performed using norm-conserving pseudopotentials 
\cite{kleinman-82-pp, troullier-91-pp}, 150 and 600 Ry kinetic energy 
cutoffs for wavefunctions and charge density, 4$\times$4$\times$4 
grids for el-ph calculations, 32$\times$32$\times$32 
grids for band structure and transport 
calculations, and 0.5 eV energy grid spacing.

For a single HH compound, the el-ph calculation takes 
about 100 core-hours on 4$\times$4$\times$4 grids 
and 4,600 core-hours on 8$\times$8$\times$8 grids. 
By extrapolating to 16$\times$16$\times$16 grids 
the cost for el-ph calculation is expected to be 
300,000 core-hours. The band structure calculation 
takes about 3 core-hours on a 32$\times$32$\times$32 
grid and 8 core-hours on a 48$\times$48$\times$48 grid. 
For a given chemical potential of electrons and temperature, 
the CRT and EPA calculations take about $0.06$ core-hours each 
when using 4$\times$4$\times$4 and 32$\times$32$\times$32 grids, 
and about $0.15$ core-hours each when using 8$\times$8$\times$8 
and 48$\times$48$\times$48 grids. In contrast, the EPW calculation 
takes about 2,600 core-hours when using 4$\times$4$\times$4 
and 32$\times$32$\times$32 grids. It is clear that direct 
BZ sampling is prohibitively expensive, and that the EPA 
approximation offers a practical alternative, much faster and 
more automatic than other interpolation approaches such as EPW.

To quantify the applicability range of the EPA approximation, 
we compute the highest optical phonon energy $\omega_\mathrm{max}$ 
for all HH compounds studied in this work. The resulting 
value of 43 meV for $\omega_\mathrm{max}$ is lower 
than 200 meV energy grid spacing (Supplementary Fig.~1). 
The Debye temperature corresponding to $\omega_\mathrm{max}$ 
($230^\circ$C) is also lower than the temperature 
at the hot side of the device used for automotive 
TE power generation ($400^\circ$C). This justifies 
the application of the EPA approximation to HH compounds.

\subsection*{Acknowledgements}
The authors thank Prof. Cheol-Hwan Park at Seoul National 
University, South Korea for his help with the \texttt{EPW} code. 
This work was supported by the U.S. Department of Energy 
under the award $\mathrm{DE}$-$\mathrm{EE0004840}$.


%

\end{document}


\title{Supplementary Information: Accelerated screening of thermoelectric materials by first-principles computations of electron-phonon scattering}

\author{Georgy Samsonidze}
\affiliation{Research and Technology Center, Robert Bosch LLC, Cambridge, MA, USA}
\author{Boris Kozinsky}
\affiliation{Research and Technology Center, Robert Bosch LLC, Cambridge, MA, USA}

\date{\today}
\maketitle

\onecolumngrid

\begin{figure}[h!]
\includegraphics[width=0.54\textwidth]{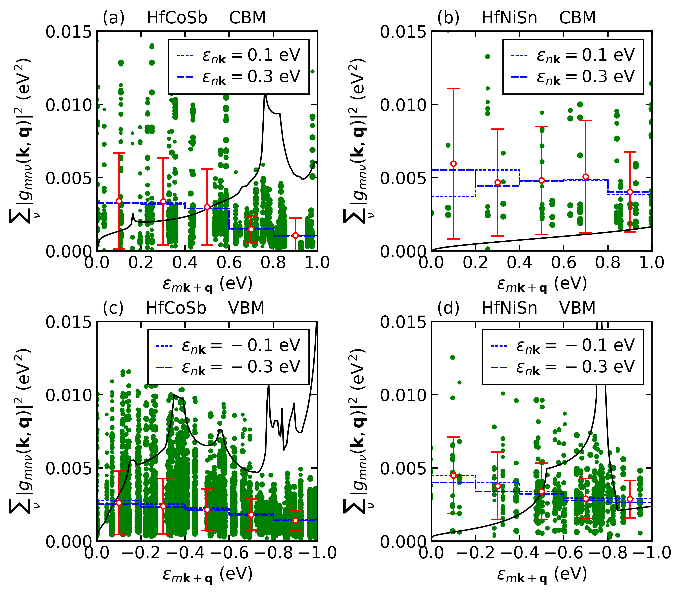}
\caption{The el-ph coupling matrix elements squared 
and summed over the phonon mode index $\mode$ 
as a function of the scattered electron energy 
$\epsilon_{\bandb\mathbf{k+q}}$ near the valence 
band maximum (VBM) and the conduction band 
minimum (CBM) of half-Heusler compounds 
\mbox{HfCoSb} and \mbox{HfNiSn}. 
Calculations are performed using a uniform 
8$\times$8$\times$8 $\Gamma$-centered grid 
for both $\mathbf{k}$- and $\mathbf{q}$-points. 
The filled green dots are the calculated 
values in the 0.1--0.3 eV range of the incident 
electron energy $\epsilon_{\banda\mathbf{k}}$. 
The dot diameters represent the relative 
weights of $\mathbf{k}$- and $\mathbf{q}$-points. 
The open red dots and error bars are the mean 
and standard deviation values over the cells 
of an energy grid with a spacing of 0.2 eV. 
The short- and long-dashed blue curves are the 
averaged values on the right-hand side of Eq.~(9) 
at $\epsilon_{\banda\mathbf{k}} = 0.1$ and $0.3$ 
eV, respectively. The averaging is performed 
over the cells of $\epsilon_{\banda\mathbf{k}}$ 
and $\epsilon_{\bandb\mathbf{k+q}}$ grids 
with the same 0.2 eV spacing. The solid 
black curve is the electron density of states 
in units of 15\AA$^{-3}$eV$^{-1}$ calculated 
using a uniform 48$\times$48$\times$48 
$\Gamma$-centered $\mathbf{k}$-point grid. 
All energies are relative to the VBM and CBM.
\label{fig:gepair}}
\end{figure}

\begin{figure}[h!]
\center{\includegraphics[width=0.27\textwidth]{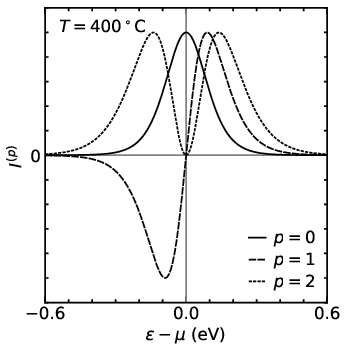}}
\caption{The solid, long-, and short-dashed curves 
show the integrand factors computed from Eq.~(6) 
for $p = 0$, $1$, and $2$, respectively, as a function 
of the electron energy $\epsilon$ relative to the chemical 
potential of electrons $\mu$ at temperature $T = 400^\circ$C. 
Vertical scaling is not identical for different curves.
\label{fig:int}}
\end{figure}